\documentclass[%
 onecolumn,
 amsmath,amssymb,
 aps,
]{revtex4-2}

\usepackage{graphicx}% Include figure files
\usepackage{dcolumn}% Align table columns on decimal point
\usepackage{bm}% bold math
\usepackage[utf8]{inputenc}
\usepackage{graphicx}
\usepackage{dcolumn}
\usepackage{bm}
\usepackage[normalem]{ulem}
\usepackage[title]{appendix}
\usepackage{comment}
\usepackage{float}
\usepackage{longtable}
\usepackage[table]{xcolor}
\usepackage{geometry}
\usepackage{amsmath}
\usepackage{amssymb}
\usepackage{natbib}

\usepackage{bm}% bold math
\usepackage{float}
\usepackage{bm}% bold math
\def \bF {\pmb{F}}
\def \bH {\pmb{H}}

\def \bx {\pmb{x}}

\def \bv {\pmb{v}}

\begin{document}

\title{Comment on ``Synchronization dynamics in non-normal networks: the
trade-off for optimality''}
\author{Francesco Sorrentino}
\author{Chad Nathe}
\affiliation{University of New Mexico, Albuquerque, US}
\date{January 2022}

\begin{abstract}
We review some of the recent literature, including Refs.\ \cite{muolo2021synchronization,nishikawa2021comment,muolo2021reply},  on the effects of non-normality on the synchronization of networks of oscillators, and provide numerical evidence that the basin of attraction about the synchronous solution is typically smaller for networks with non-normal Laplacian matrix, compared to networks with a normal Laplacian matrix.
\end{abstract}

\maketitle

Reference \cite{muolo2021synchronization} has been harshly criticized in \cite{nishikawa2021comment}. However, \cite{muolo2021synchronization} is one of the first papers to highlight the important connection between synchronization of networks of oscillators and non-normality of the network coupling matrix. Other works that have addressed the role of non-normality on nonlinear dynamics are \cite{trefethen1993hydrodynamic, trefethen2020spectra}  and its effects on networks are \cite{ravoori2011robustness,fish2017construction,nicolaou2020non}. 

The general framework to study network synchronization on which Ref.\ \cite{muolo2021synchronization} is based is that of the master stability function (MSF) \cite{Pe:Ca}, which involves the following steps:  (1) linearization of the nonlinear equations; (2) decoupling of the linearized equations in independent modes; and (3) calculation of the maximum Lyapunov exponents corresponding to each mode. It is well known that the validity of the MSF approach is limited to infinitesimal perturbations about the synchronous solution (local stability.) However, in all applications, one is interested in the effects of finite perturbations. If the synchronous solution is locally stable, it becomes important to characterize the basin of attraction, i.e., the set of initial conditions that converge to the synchronous state. The MSF approach does not characterize the basin of attraction, which instead can be done by using other methods such as those based on the analysis of Lyapunov functions \cite{Wang:Chen02} and the basin method \cite{menck2013basin}. %This has not been considered in \cite{muolo2021synchronization}, which instead remains focused on the study of linear stability.

For the case of complete synchronization, the equations describing the time evolution of the network systems are generally written as follows,
\begin{equation}
\begin{array}{cc}
\dot{\bx}_i(t)=\bF(\bx_i(t))+\sigma \sum_j A_{ij} [\bH(\bx_j(t))-\bH(\bx_i(t))], \quad i=1,...,N,
\end{array}
\end{equation}
where ${\bx}_i(t) \in R^m$ is the state of oscillator $i$, the nonlinear function $\bF:R^m \rightarrow R^m$ describes the time evolution of an uncoupled system, the nonlinear function $\bH:R^m \rightarrow R^m$ describes how one oscillator interacts with another one, $\sigma$ is the tunable coupling strength, and  the adjacency matrix $A=\{A_{ij}\}$ describes the network connectivity, i.e., $A_{ij}\geq 0$ is the strength of the coupling from node $j$ to node $i$ ($A_{ij}= 0$ if there is no coupling from node $j$ to node $i$.) Eq.\ (1) can be rewritten as follows,
\begin{equation}
\begin{array}{cc}
\dot{\bx}_i(t)=\bF(\bx_i(t))+\sigma \sum_j L_{ij} \bH(\bx_j(t)), \quad i=1,...,N, \label{Lapl}
\end{array}
\end{equation}
where the Laplacian matrix $L=\{L_{ij}=A_{ij}-\delta_{ij}\sum_j A_{ij} \}$ has sums of the entries over its rows equal to zero, $\sum_j L_{ij}=0$. Eq.\ \eqref{Lapl} allows a complete synchronous solution $\bx_1(t)=\bx_2(t)=...=\bx_N(t)=\bx_s(t)$, which obeys the equation of an uncoupled system $\dot{{\bx}}_s(t)=\bF({\bx}_s(t))$.

We now write the eigenvalue equation for the Laplacian matrix
\begin{equation}
    L \bv_k =\lambda_k \bv_k,
\end{equation}
and note that by the property that the sums of the rows of the matrix $L$ are equal to zero, at least one of the eigenvalues $\lambda_1=0$. Under the assumption that the network is weakly
connected, $\Re(\lambda_k)<0$, $k=2,...,N$. Moreover, if the spectrum is real the eigevalues can be ordered so that $\lambda_1 > \lambda_2 \geq ... \geq \lambda_N$.

%For the sake of brevity, we skip the derivations, which 
In order to study local stability, the set of equations \eqref{Lapl} are linearized about the synchronous solution and decoupled in the eigenvalues of the Laplacian matrix, which for the moment we assume to be diagonalizable.
The detailed derivations can be found in \cite{Pe:Ca}. Here we present the final result that one can introduce a master stability function which associates a maximum Lyapunov exponent $\mathcal{M}(\sigma \lambda_k)$ to each $a=\sigma \lambda_k$, $k=1,...,N$. The condition for local stability is that 
\begin{equation} \label{condition}
\mathcal{M}(\sigma \lambda_k)<0, 
\end{equation}
$k=2,...,N$. 
The advantage of this approach is that one can pre-compute the values of $a$ for which $\mathcal{M}(a)<0$ and then verify whether all $\sigma \lambda_k$, $k=2,..,N$ fall in the region of the complex plane where $\mathcal{M}(a)<0$.
It is important to emphasize that condition \eqref{condition} only depends on the eigenvalues $\lambda_2,\lambda_3,...,\lambda_N$ and does not depend on the eigenvectors.

Other research has pointed out the importance of the eigenvectors of the Laplacian. In particular, in the case in which the matrix $L$ has repeated eigenvalues and is non-diagonalizable, the MSF theory needs to be adjusted so that the Laplacian matrix is diagonalized in Jordan blocks \cite{nishikawa2006synchronization}. Then a Lyapunov exponent is associated to each Jordan block, instead of each eigenvalue. However, Reference \cite{nishikawa2006synchronization} has shown that the condition for local stability of the synchronous solution is still  given by \eqref{condition} with the caveat that when the eigenvalues are coupled through a Jordan blocks a longer transient to the synchronous state can be seen, compared to the case in which the equations are decoupled in scalar blocks. References \cite{nishikawa2006synchronization,nishikawa2010network,ravoori2011robustness} studied the case that the matrix $L$ has repeated eigenvalues and is non-diagonalizable. We call $\mathcal{D}$ the set of matrices that are nondiagonalizable.

A separate relevant observation is that the range of the coupling $\sigma$ over which the synchronous solution is stable (the so called `synchronizability' \cite{report}) is maximized when the eigenvalues $\lambda_2=\lambda_3=...\lambda_N$. Following \cite{nishikawa2006synchronization,nishikawa2010network,ravoori2011robustness}, in what follows we call these networks \textit{optimal}. So on the one hand, having repeated eigenvalues is good  because it maximizes the synchronizability, on the other hand it's bad because it may slow down the convergence to the synchronous state \cite{nishikawa2006synchronization}.

Reference \cite{muolo2021synchronization} has pointed another effect of the eigenvectors in the case that the matrix $L$ is non-normal, i.e., $L L^T \neq L^T L$.  All symmetric matrices are normal, while asymmetric matrices are generically non-normal, with some exceptions. The issue is that the validity of the master stability function analysis is limited to infinitesimal perturbations (local stability). We call $\mathcal{N}$ the set of matrices that are nonnormal and note that $\mathcal{N} \supset \mathcal{D}$. Local stability is a necessary prerequisite for nonlocal stability, which describes convergence to the synchronous solution in the presence of finite perturbations, which is the case of actual interest in applications. If the perturbations are finite, then it is possible that a finite perturbation will exit the basin of attraction of the locally stable synchronous solution so to prevent convergence on the synchronous solution. %In this case the transient dynamics also plays a role. 
The claim of Refs.\ \cite{muolo2021synchronization,muolo2021reply} is that \textit{the presence of non-normal eigenvectors will reduce the finite basin of attraction.}

To conclude, in general both the eigenvalues and the eigenvetors of the Laplacian affect the synchronization of networks described by Eq.\ \eqref{Lapl}. The eigenvalues determine local stability of the synchronous solution (both in the case of diagonalizable and non-diagonalizable Laplacian matrices), while the eigenvectors may affect the basin of attraction of the synchronous solution (nonloncal stability.)  An important difference between \cite{nishikawa2006synchronization} and \cite{muolo2021synchronization} is that \cite{nishikawa2006synchronization} only focused on linear stability and reported a longer transient towards the synchronous solution in the non-generic case that the Laplacian matrix is non-diagonalizable, $L \in \mathcal{D}$, while \cite{muolo2021synchronization} dealt with the generic case that the Laplacian matrix is nonnormal, $L \in \mathcal{N} \supset D$.

Here we test the claim presented in  \cite{muolo2021synchronization,muolo2021reply}  is simulation. The issue with using different Laplacian matrices is that they may differ in both their eigenvalues and eigenvectors (see eg. the example in Fig.\ 1 of \cite{muolo2021synchronization}.) Therefore, in this note we fix the eigenvalues and vary the eigenvectors. Then test the claim that non-normality may affect the basin of attraction about the synchronous solution. We emphasize that an important step of our numerical analysis is that different from \cite{muolo2021synchronization}, we compare networks with the same set of eigenvalues. Therefore, these networks have the same (negative) maximum Laypunov exponents  that describe the asymptotic rate of convergence towards the synchronous solution. Varying the eigenvectors while having fixed the eigenvalues and the coupling strength $\sigma$ allows us to fairly compare the role of the eigenvectors in determining the basin of attraction for different networks.

We take the network equations to be described by Eq. (2) and choose the initial conditions to be,
    \begin{equation}
        \textbf{x}_i(0) = \textbf{x}_0+\textbf{r}D,
    \end{equation}
    $i=1,...,N$, where $\textbf{x}_0$ is a randomly chosen point from the attractor of an uncoupled oscillator, the vector $\textbf{r}$ has entries that are randomly drawn from a {normal distribution} and the scalar $D$ is the magnitude of the perturbation. For a given value of $D$ we produce different realizations of $\textbf{x}_i(0)$ for $i=1,..,N$ and average the results. For $D$ small enough (large enough) we expect the network to synchronize (not synchronize), which is measured in terms of the synchronization error,
     \begin{equation}\label{err}
        \text{Error} = N^{-1} \left< \left<  \sum_{i=1}^N x_i(t) - \overline{\mathbf{x}(t)} \right>_t \right>_0,
    \end{equation}
    where  the notation, $\overline{\mathbf{x}(t_f)}$ indicates an average over the oscillators, the notation $\left< \right>_t$ indicates a time average for large times $t$, the notation $\left< \right>_0$ indicates an average over different choices of the initial conditions. As the synchronization error becomes positive for $D>D_c$, we take $D_c$ as a rough measure of the basin of attraction. %grows past a certain value, it will eventually leave the basin of attraction, and lead to a positive synchronization error

%\begin{equation}\label{gen}
%    \mathbf{\dot{x}}_i(t) = \mathbf{F}(\mathbf{x}_i(t))+\sigma\sum_{j=1}^N L_{ij} [\mathbf{H}(\mathbf{x}_i(t))]
%\end{equation}
%$i=1,...,N$,
%where $\mathbf{{x}}_i(t)  \in \R^n$ is the state of oscillator $i$ at time $t$, 
 %    $\mathbf{{F}}: \R^n\mapsto \R^n$ indicates the dynamics of each uncoupled node and $\mathbf{{H}}: \R^n\mapsto \R^n$ indicates the coupling between nodes. The matrix $L=\{L_{ij} \}$ represents the network connectivity; we call $\mathbf{H}(\mathbf{x}_j(t))$ the \textit{signal} received at node $i$ from node $j$, and the full graph, $L=\{L_{ij}\}$. We pick $\sigma$ such that all the nodes in the network will synchronize given that the initial conditions are in the basin of attraction of the oscillators. We determine $\sigma$ to be,
     
     %\begin{equation}
     %    \sigma = \frac{-K}{\Lambda}
     %\end{equation}
    %where $\Lambda$ is the largest real eigenvalue of $A$. With a large enough $K$, the network will become stable. We 

    The three networks we consider are shown in Fig.\ 1. The network in panel (A) is fully connected with uniform weights, $A_{ij}=1/N-\delta_{ij} 1/N$. The networks in panel (B) is the unidirectional star graph, $A_{ij}=1$ for $j=1$ and $i\neq 1$, $A_{ij}=0$ otherwise. The network in panel (C) is the  unidirectional chain graph  $A_{ij}=1$ for $i=2,...,N$ and $j=(i-1)$, $A_{ij}=0$ otherwise.
    Following the terminology in \cite{nishikawa2006synchronization,nishikawa2010network,ravoori2011robustness}, we call the network in (A)
    an optimal undirected network, the network in (B) an optimal directed non-sensitive network and the network in (C)
    an optimal directed sensistive  network. They are all `optimal' as they have real spectrum and the same set of eigenvalues $\lambda_1=0$, $\lambda_2=\lambda_3=...=\lambda_N=-1$. However, the eigenvectors are different from network to network. In particular, the Laplacian matrix of the fully connected graph is normal, while the Laplacian matrices of the unidirectional star network and that of the unidirectional chain network are non-normal. 

%We introduce two new networks. One where the network is unidirectional where the first node drives the second, and the second drives the third and so on. The second, is where the first node drives all the nodes in the network, and there are no other connections. We refer to these networks as the chain and the star respectively. 
\begin{figure}[H]
\centering
    \begin{tabular}{l l l}
       \text{(A)} & \text{(B)} & \text{(C)}\\
\includegraphics[width=.3\textwidth]{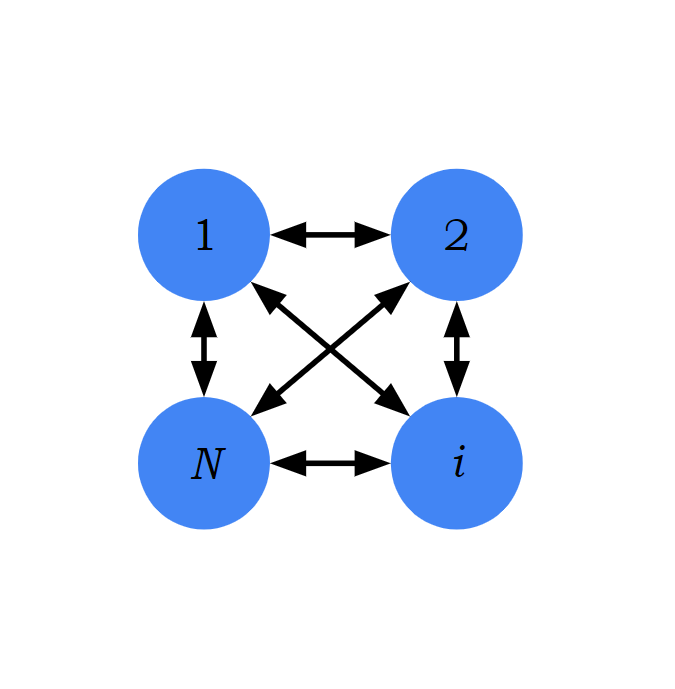}  &
\includegraphics[width=.3\textwidth]{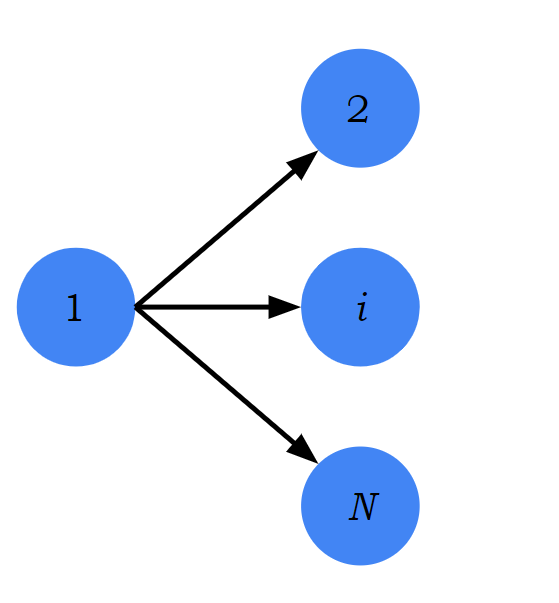} &
\includegraphics[width=.4\textwidth]{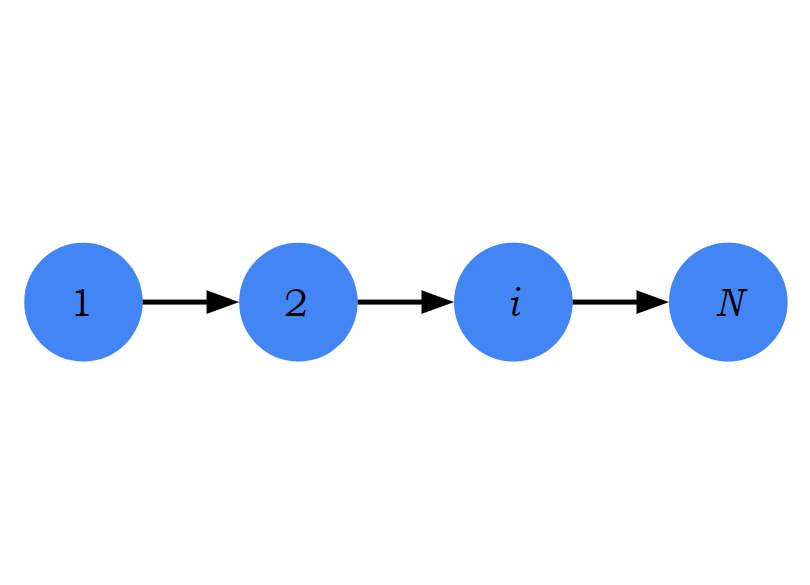} 
   \end{tabular}
    \caption{(A) $L_{orth}$ network; (B) Unidirectional Star network; (C) Unidirectional Chain Network.  Here, the number of nodes $N$ is variable.} 
\end{figure}

The results of our numerical simulations are shown in Fig.\ 2. All networks studied in Fig.\ 2 have $N=30$ nodes.  Each panel of the figure refers to a different choice of the functions $\bF$ and $\bH$, namely, panel (A) is for the Chen system, coupled in the $y$-variable, $\sigma=60$; panel (B) is for Chua systems, coupled in the $x$-variable, $\sigma=30$; panel (C) is for Van der Pol systems, coupled in the $y$-variable, $\sigma=1$; panel (D) is for Duffing systems, $\sigma=1$, coupled in the $x$-variable. Details about each case are provided in the Appendix. 

In all our simulations we always find the critical $D^c$ for the fully connected graph  to be larger than the critical $D^c$ for the unidirectional star graph, and the latter to be much larger (by several orders of magnitude) than the critical $D^c$ for the unidirectional chain graph. This indicates that the basin of attractions are significantly different from network to network and confirms the claim of Refs. \cite{muolo2021synchronization,muolo2021reply}. In particular, the chain network with non-diagonalizable Laplacian, though considered optimal, is characterized by an appreciably smaller basin of attraction than the other two network (see in particular panels (B) and (D)).

\begin{figure}[H] 
\centering
    \begin{tabular}{l l}
       \text{(A)} & \text{(B)}\\
\includegraphics[width=.45\textwidth]{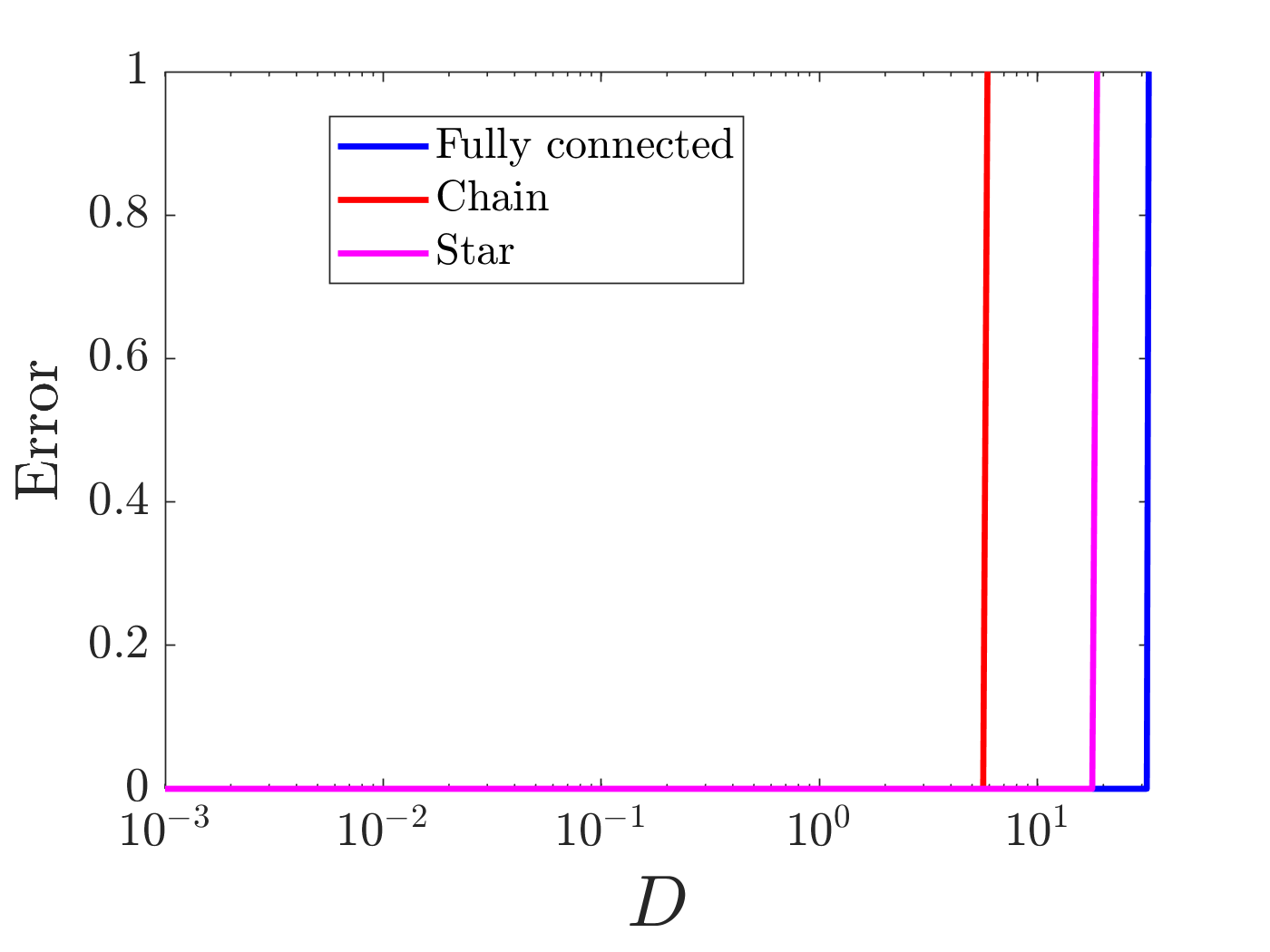} &
\includegraphics[width=.45\textwidth]{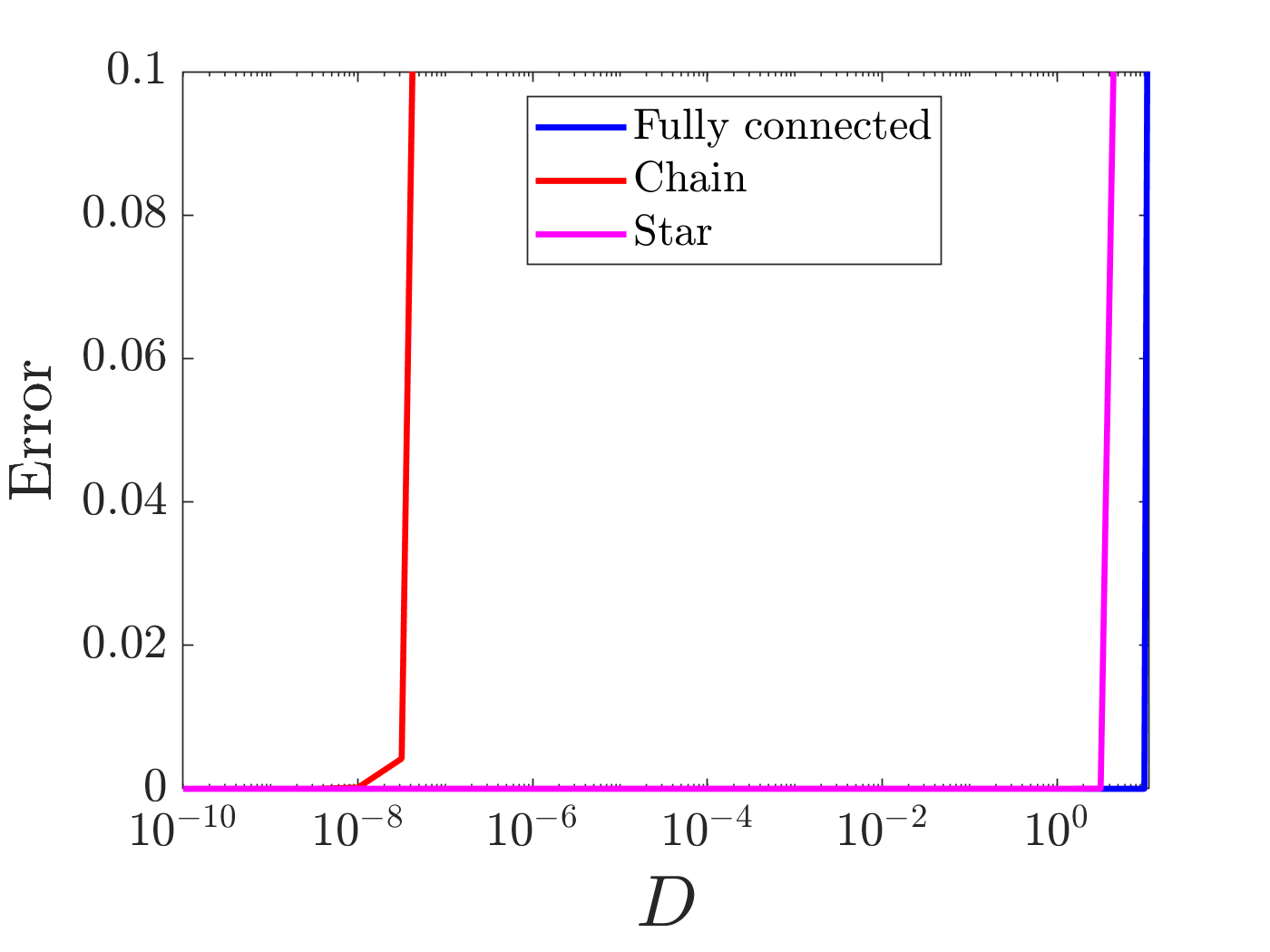}
\end{tabular}
\begin{tabular}{l l}
       \text{(C)} & \text{(D)}\\
\includegraphics[width=.45\textwidth]{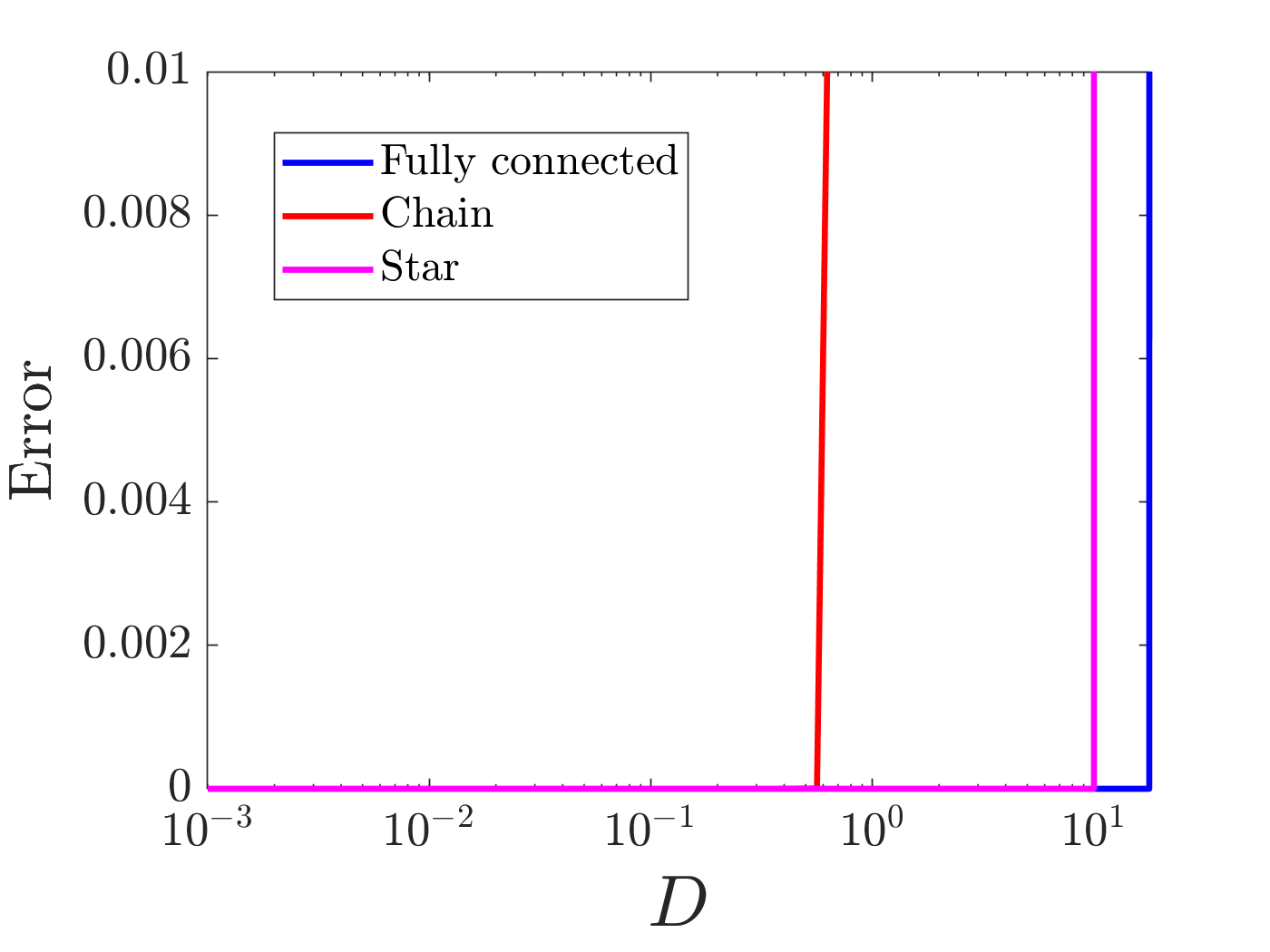} &
\includegraphics[width=.45\textwidth]{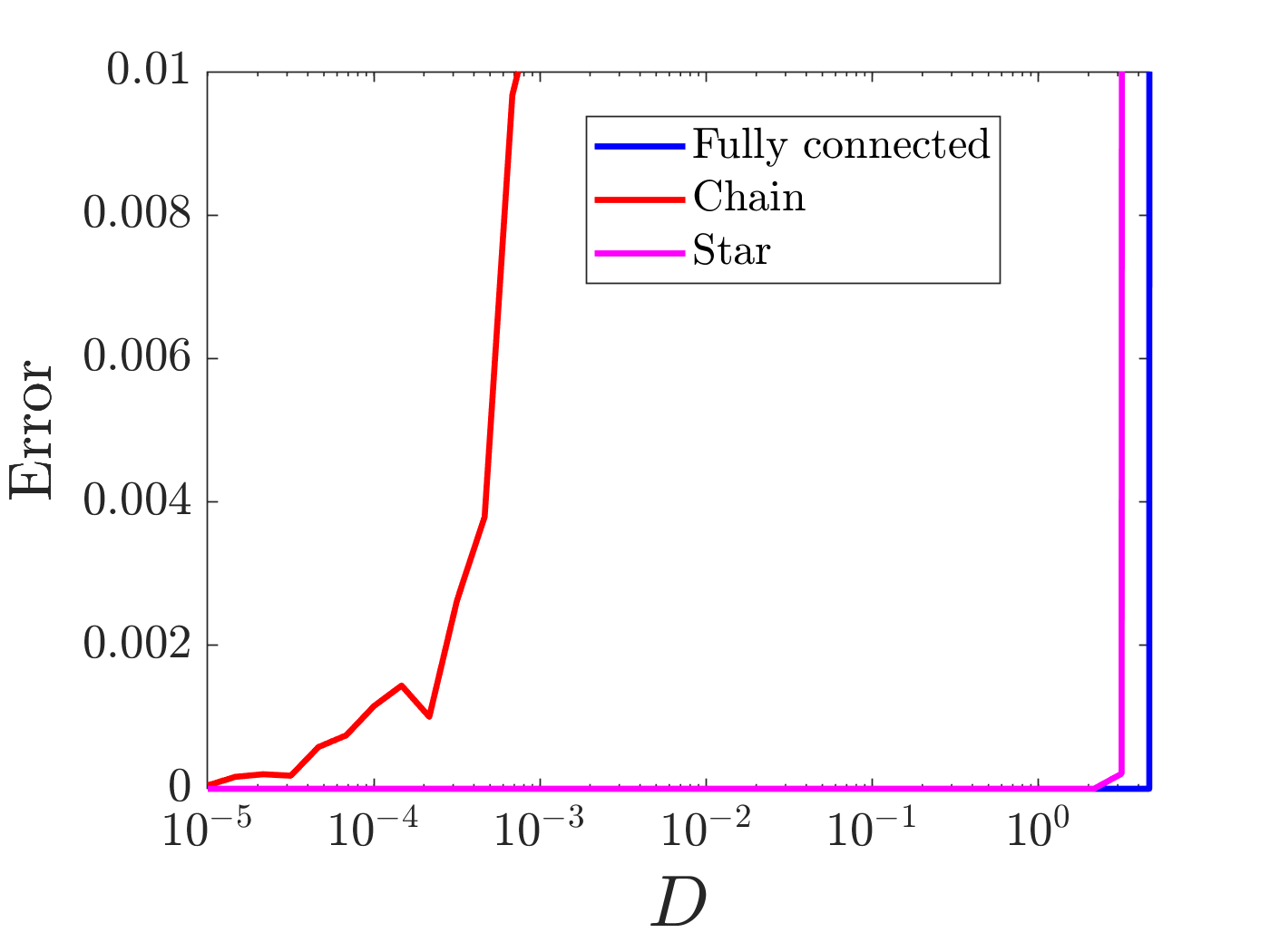}
\end{tabular}
\caption{Comparison of the synchronization error (Eq.\ \eqref{err}) for three networks with $N=30$ nodes: fully connected network (in blue), unidirectional chain (in red), and unidirectional star (in magenta). Each panel shows a different choice of the functions $\bF$ and $\bH$: (A) Chen systems coupled in the $y$-variable, $\sigma=60$; (B) Chua systems coupled in the $x$ variable, $\sigma=30$; (C) Van der Pol systems coupled in the $y$ variable, $\sigma=1$; (D) Duffing systems coupled in the $x$ variable, $\sigma=1$.} 
\end{figure}

We extend our analysis of the basin of attraction to a comparison between an optimal chain network and a non-optimal network. We show our results for both Chua systems coupled in the $x$-variable and Van der Pol systems coupled in the $y$-variable, but obtained similar results for the Chen system and the Duffing system (see Appendix.) The optimal chain network is the network in Fig.\ 1(C) with $N=30$ nodes. The non-optimal network is a connected undirected Erdos-Renyi graph with $N=30$ nodes and four edges per node. The Laplacian matrix for this network is normalized  $L \rightarrow -{\lambda_2}^{-1}\times {L}$ so that the largest nonzero eigenvalue is equal to $-1$, the same as the chain network. The rational for this normalization is that the MSF for the  Chua systems coupled in the $x$-variable and for the Van der Pol systems coupled in the $y$-variable is negative in the unbounded real interval $(-\infty,a_{min})$, hence if $\sigma \lambda_2< a_{min}$, so are $\sigma \lambda_3, \sigma \lambda_4,..., \sigma \lambda_N$. Thus  both networks are stable for $\sigma>|a_{min}|$. Our results are shown in Fig.\ 3 where the synchronization error  is plotted vs $\mbox{Log}_{10}(D)$ and $\sigma>|a_{min}|$. As can be seen, for all values of $\sigma$ shown, the basin of attraction for the random network is much larger than the basin of attraction for the chain network. Our results are not in contrast with those of Ref.\ \cite{nishikawa2006synchronization} but highlight that the size of the basin of attraction is not captured by the traditional measures of network synchronizability \cite{Ba:Pe02}. % Our results partially contrast with those of \cite{nishikawa2010network} because in that paper

\begin{figure}[H] 
\centering
    \begin{tabular}{l l}
       \text{(A)} & \text{(B)}\\
\includegraphics[width=.49\textwidth]{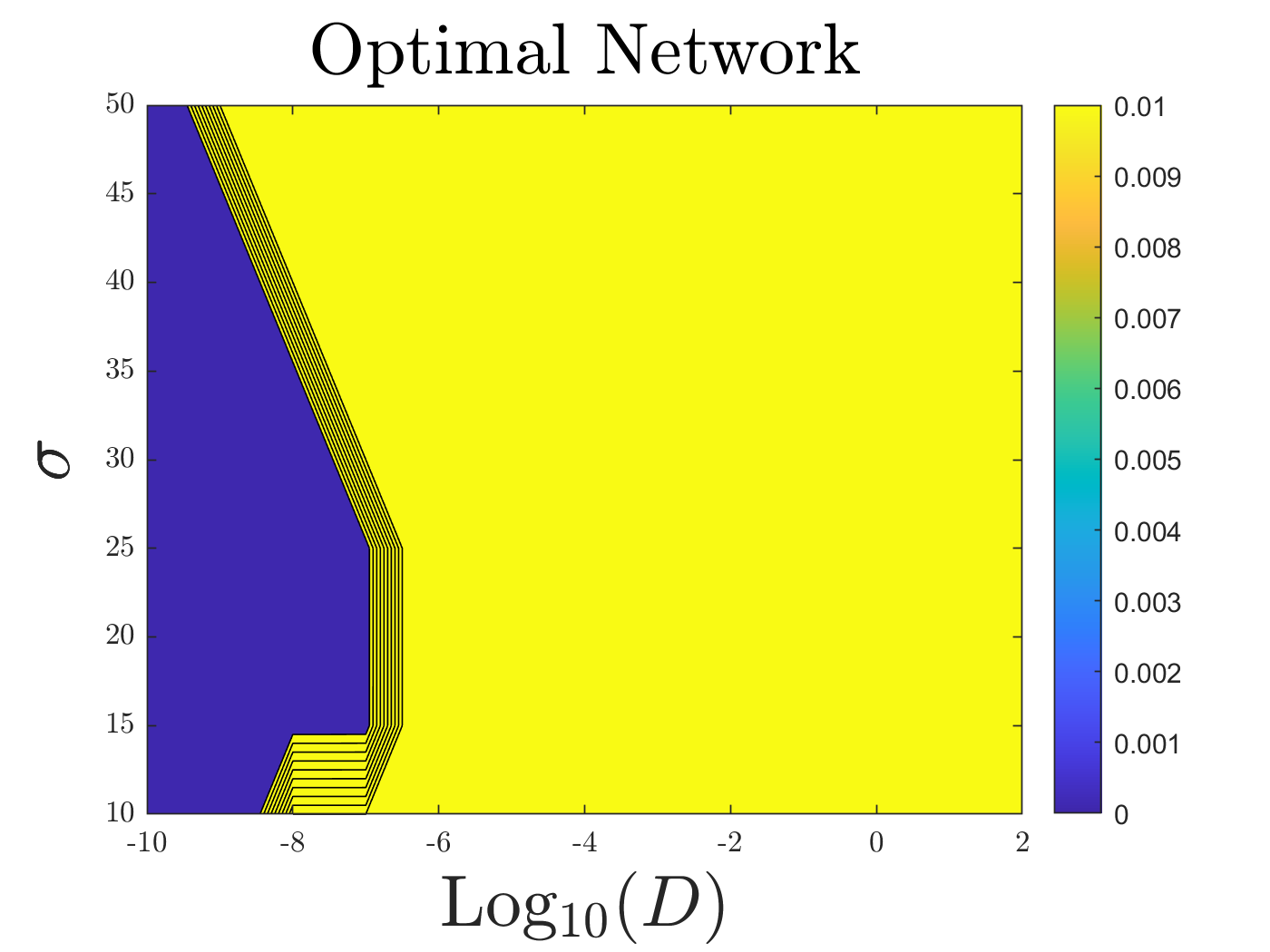} &
\includegraphics[width=.49\textwidth]{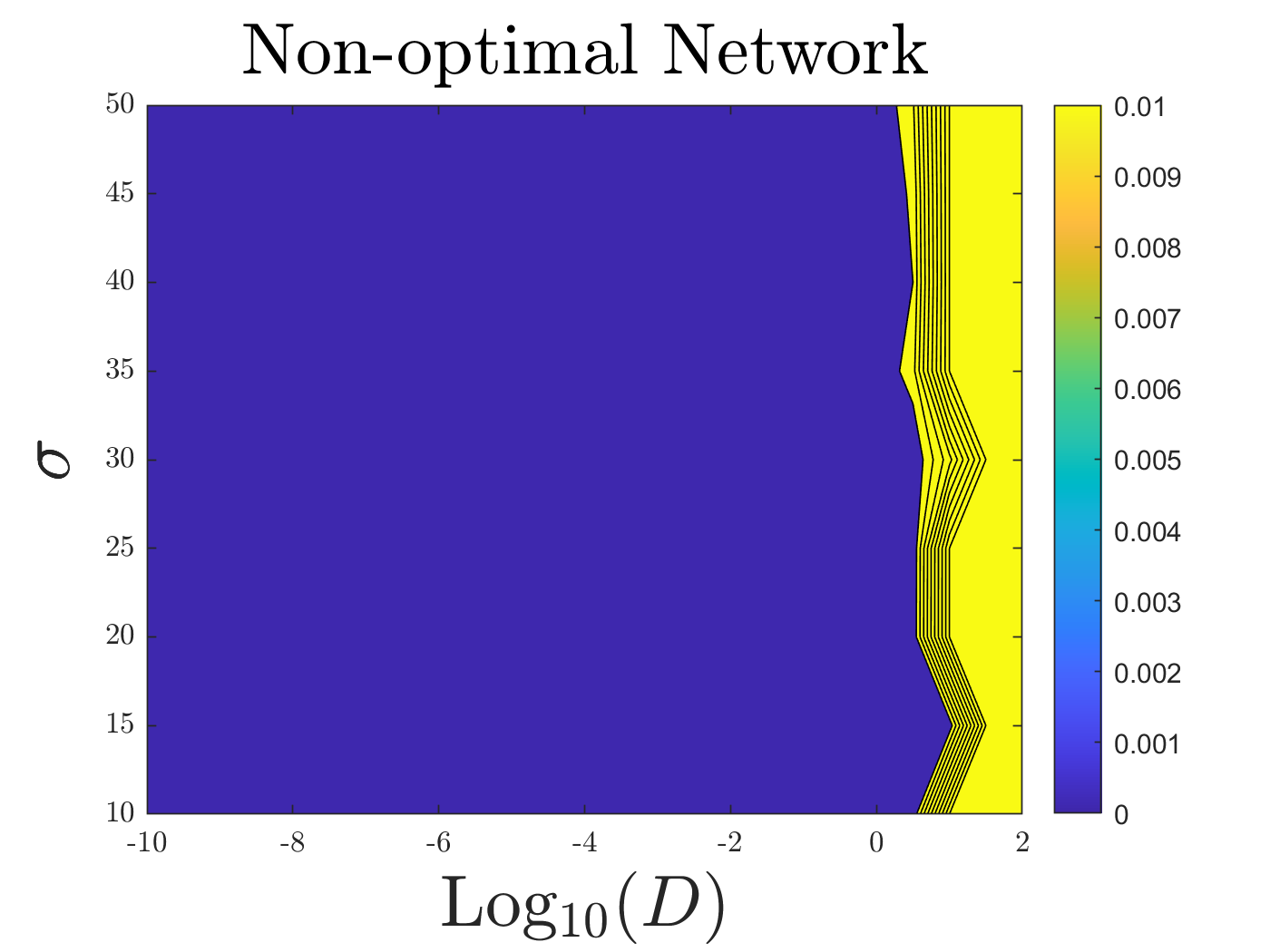}
\end{tabular}
  \begin{tabular}{l l}
       \text{(A)} & \text{(B)}\\
\includegraphics[width=.49\textwidth]{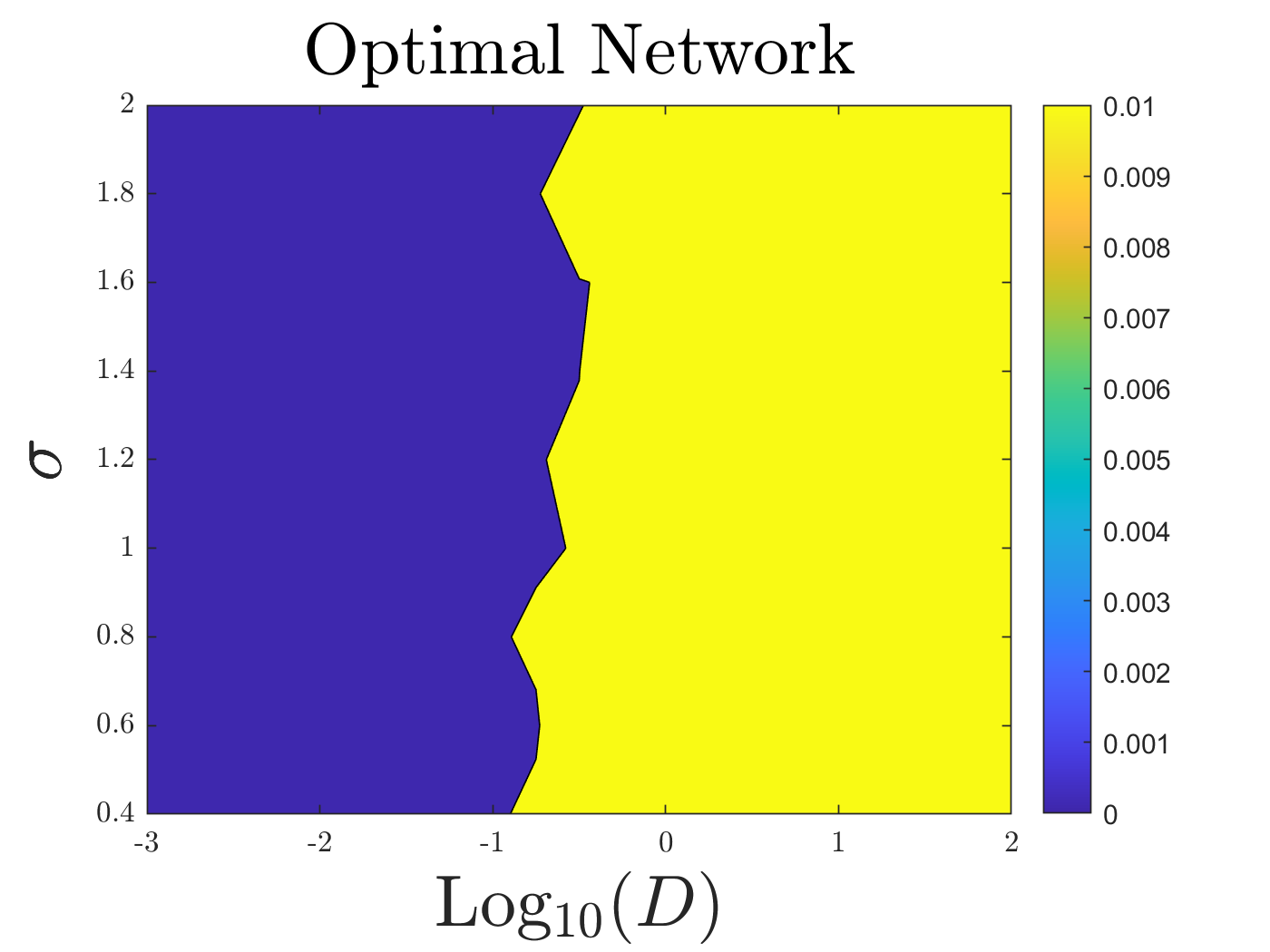} &
\includegraphics[width=.49\textwidth]{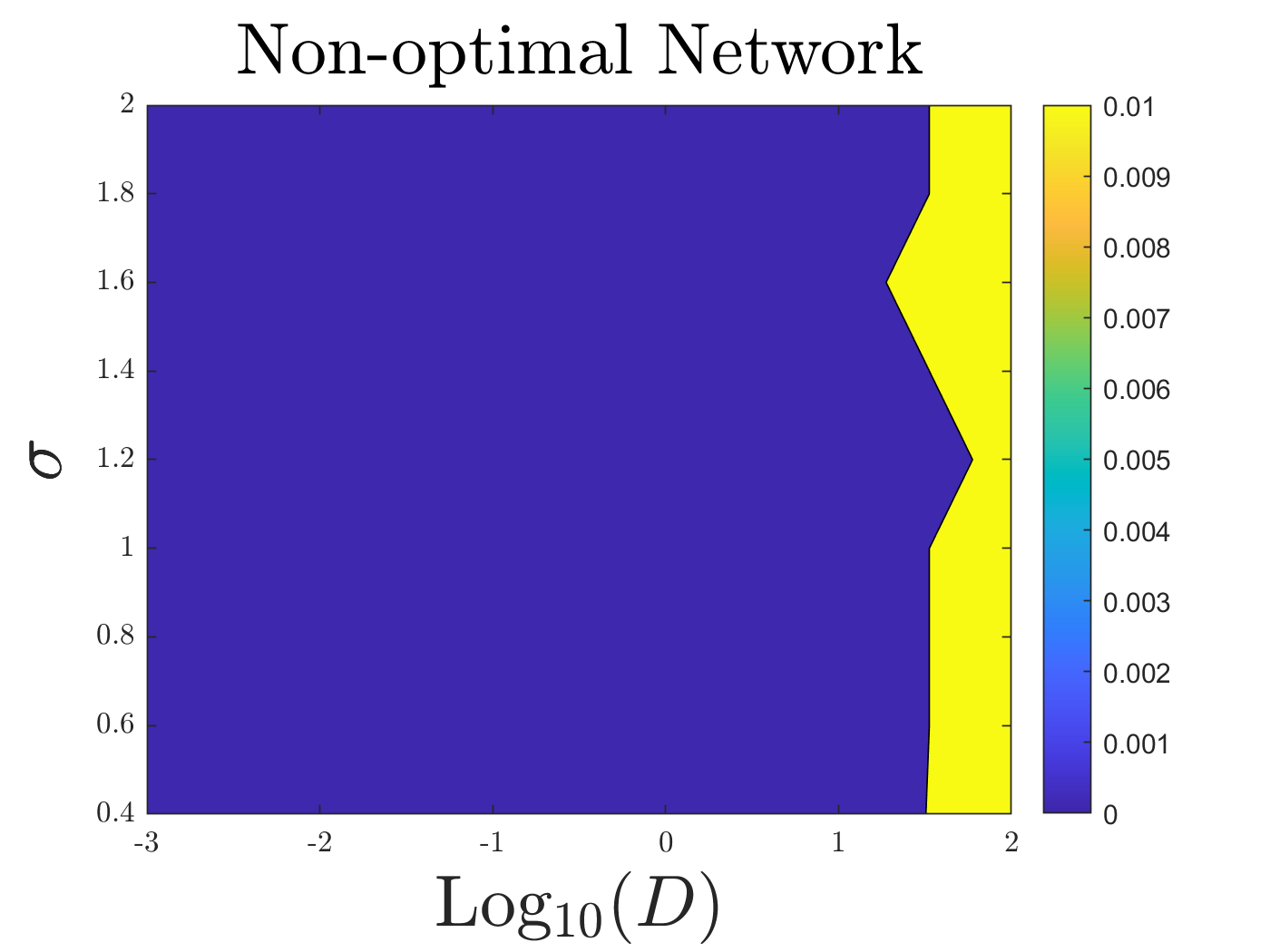}
\end{tabular}
\caption{The synchronization error (Eq.\ \eqref{err}) is plotted vs $\mbox{Log}_{10}(D)$ and $\sigma$. (A) Chua systems coupled in the $x$-variable and optimal chain network. (B) Chua systems coupled in the $x$-variable and non-optimal random network. (C) Van der Pol systems coupled in the $y$-variable and optimal chain network. (D) Van der Pol systems coupled in the $y$-variable and non-optimal random network.} 
\end{figure}

%\begin{figure}[H] 
%\centering
%
%\includegraphics[width=.4\textwidth]{Figures/chainstar.jpg}
%\includegraphics[width=.4\textwidth]{Figures/duffingchainstar.jpg}
%\caption{Error versus $D$. We compare three networks, the star, the chain and $L_{orth}$. $N=30$. Chua system.} 
%\end{figure}

To conclude we have tested numerically the connection between synchronization of networks of oscillators and non-normality of the network coupling matrix.
Our numerical results confirm some of the claims presented in  Refs.\ \cite{muolo2021synchronization,muolo2021reply} and indicate  that  networks with non-normal Laplacian matrix are typically characterized by a smaller basin
of attraction than networks with a normal Laplacian matrix. Our work also complements some of the early results in \cite{fish2017construction}.  We expect similar considerations to also apply to the case of cluster synchronization \cite{lodi2021one}. Finally, our work points out the need for an alternative measure of network synchronizability %, %possibly based on the eigenvectors of the Laplacian matrix \cite{muolo2021synchronization,muolo2021reply}, 
which addresses the important issue of the basin of attraction for the synchronous solution.

\section*{Appendix}
Our numerical simulations are based on four different choices of the functions $\bF$ and $\bH$, which we describe below.

    \centerline{\it{\textbf{1. Chen Oscillators coupled in the $y$ variable}}}

\begin{equation}
    \begin{aligned}
    & \dot x = a(y-x)\\
    & \dot y = (c -a -z)x + cy  \\ %+ \sigma A y\\
    & \dot z = xy -\beta z
    \end{aligned}
\end{equation}
\\
where $a=35$, $c=28$, $\beta = \frac{8}{3}.$ We couple the Chen oscillators in the $y$-variable, i.e.,
\begin{equation}
    \bH(\bx)=\left[ {\begin{array}{ccc}
   0 & 0 & 0 \\
   0 & 1 & 0 \\
   0 & 0 & 0
  \end{array} } \right] \bx
\end{equation}

\centerline{\it{\textbf{2. Chua Oscillators coupled in the $x$ variable}}}

\begin{equation}
    \begin{aligned}
    & \dot x = \alpha[y -x + f(x)]  \\%+ \sigma A y\\
    & \dot y = x -y + z\\
    & \dot z = \beta y - \gamma z
    \end{aligned}
\end{equation}
\\
where $\alpha = 10$, $\beta = 14.87$, $\gamma = 0$, and
\\
\begin{equation}
f(x) = 
 \begin{cases} 
      -bx -a + b, & x > 0 \\
      -ax, & |x| < 1 \\
      -bx + a -b & x < -1.
   \end{cases}
\end{equation}
We couple the Chua oscillators in the $x$-variable, i.e.,
\begin{equation}
    \bH(\bx)=\left[ {\begin{array}{ccc}
   1 & 0 & 0 \\
   0 & 0 & 0 \\
   0 & 0 & 0
  \end{array} } \right] \bx
\end{equation}
\\
\centerline{\it{\textbf{3. Van der Pol Oscillators coupled in the $y$ variable}}}

\begin{equation}
    \begin{aligned}
    & \dot x = y  \\% + \sigma A x\\
    & \dot y = -x +d(1-x^2)y + Z\sin(\eta t)  + \sigma A y
    \end{aligned}
\end{equation}
\\
where $d=3$, $Z=15$, $\eta = 4.065$. We couple the Van der Pol oscillators in the $y$-variable, i.e.,
\begin{equation}
    \bH(\bx)=\left[ {\begin{array}{cc}
   0 & 0 \\
   0 & 1  \\
  \end{array} } \right] \bx
\end{equation}
\\
\centerline{\it{\textbf{4.Forced Duffing Oscillators coupled in the $x$ variable}}}

\begin{equation}
    \begin{aligned}
    & \dot x = y  \\%+ \sigma A x\\
    & \dot y = -hy -x^3 + q\sin(\eta t)
\end{aligned}
\end{equation}
\\
where $\eta=1$, $h=0.1$, and $q=5.6$. We couple the Duffing oscillators in the $y$-variable, i.e.,
\begin{equation}
    \bH(\bx)=\left[ {\begin{array}{cc}
   1 & 0 \\
   0 & 0  \\
  \end{array} } \right] \bx
\end{equation}
    
% \newcommand{\noop}[1]{}

% \bibliography{lib}
%apsrev4-2.bst 2019-01-14 (MD) hand-edited version of apsrev4-1.bst
%Control: key (0)
%Control: author (8) initials jnrlst
%Control: editor formatted (1) identically to author
%Control: production of article title (0) allowed
%Control: page (0) single
%Control: year (1) truncated
%Control: production of eprint (0) enabled
 \newcommand{\noop}[1]{}

\end{document}